\documentstyle[12pt]{article}
\def\sss{\scriptscriptstyle}
\def\barp{{\raise.35ex\hbox{${\sss (}$}}---{\raise.35ex\hbox{${\sss )}$}}}
\def\bdbarp{\hbox{$B_d$\kern-1.4em\raise1.4ex\hbox{\barp}}}
\def\bsbarp{\hbox{$B_s$\kern-1.4em\raise1.4ex\hbox{\barp}}}
\newcommand{\xd}{x_d}
\newcommand{\xs}{x_s}
\newcommand{\bd}{B_d^0}

\newcommand{\bs}{B_s^0}

\newcommand{\bu}{B_u^\pm}
\newcommand{\beq}{\begin{equation}}
\newcommand{\eeq}{\end{equation}}
\newcommand{\absvcb}{\vert V_{cb}\vert}
\newcommand{\absvub}{\vert V_{ub}\vert}

\newcommand{\abseps}{\vert\epsilon\vert}

\newcommand{\fbb}{f^2_{B_d}B_{B_d}}
\newcommand{\fbbs}{f^2_{B_s}B_{B_s}}
\newcommand{\fbd}{f_{B_d}}
\newcommand{\fbs}{f_{B_s}}

\def\att{t \bar{t}}
\def\app{p \bar{p}}
\def\rts{\sqrt{s}}
\def\mt{m_t}
\def\mb{m_b}
\def\mc{m_c}

\newcommand{\delm}{\Delta M}

\newcommand{\kkbar}{$K^0$-${\overline{K^0}}$}
\newcommand{\bdbdbar}{$B_d^0$-${\overline{B_d^0}}$}
\newcommand{\bsbsbar}{$B_s^0$-${\overline{B_s^0}}$}

\begin{document}
\begin{flushright}
CERN-TH.7248/94\\
UdeM-LPN-TH-94-197\\
\end{flushright}
\begin{center}
{\Large \bf
\centerline
{Implications of the Top Quark Mass Measurement}
\vspace*{0.3cm}
\centerline
{for the CKM Parameters, {\bf $\xs$} and CP Asymmetries}}
\vspace*{1.5cm}
\vskip1cm
 {\large A.~Ali}$\footnote{On leave of absence from DESY, Hamburg, FRG.}$
\vskip0.2cm
       Theory Division, CERN  \\
       CH-1211 Geneva 23, Switzerland \\
\vspace*{0.5cm}
\centerline{ and}
\vspace*{0.5cm}
\centerline{\large D. ~London}
\smallskip
       Laboratoire de physique nucl\'eaire, Universit\'e de
Montr\'eal \\
             C.P. 6128, succ. centre-ville, Montr\'eal, QC, Canada
H3C 3J7\\
\vskip1cm
{\Large Abstract\\}
\parbox[t]{\textwidth}{
\indent
Motivated by the recent determination of the top quark mass by the CDF
collaboration, $\mt =174 \pm 10 ^{+13}_{-12}$ GeV, we review and update
constraints on the parameters of the quark flavour mixing matrix $V_{CKM}$
in the standard model. In performing these fits, we use inputs from the
measurements of $\abseps$, the CP-violating parameter in $K$ decays, $\xd =
(\delm)/\Gamma$, the mixing parameter in \bdbdbar\ mixing, the present
measurements of the matrix elements $\absvcb$ and $\absvub$, and the
$B$-hadron lifetimes. The CDF value for $\mt$ considerably reduces the
CKM-parameter space previously allowed. An interesting result of our
analysis is that the present data can be used to restrict the coupling
constant product ratio $f_{B_d}\sqrt{B_{B_d}}$ to the range 110-270 MeV --
in comfortable agreement with existing theoretical estimates of this
quantity. We use the updated CKM matrix to predict the \bsbsbar\ mixing
ratio $\xs$, as well as the quantities $\sin 2\alpha$, $\sin 2\beta$ and
$\sin^2\gamma$, which characterize
 CP-violating asymmetries in $B$-decays.
}
\vskip2cm
{\em Submitted to Physics Letters B}
\end{center}
\noindent
CERN-TH.7248/94\\
May 1994
\newpage
\setcounter{page}{1}
\textheight 23.0 true cm


\noindent
{\bf 1. Introduction}
\bigskip

The CDF collaboration at Fermilab has recently published evidence for top
quark production in $\app$ collisions at $\rts = 1.8$ TeV. The search is
based on the final states expected in the decays of the top quark in the
standard model (SM). Based on this analysis a top quark mass $\mt = 174 \pm
10 ^{+13}_{-12}$ GeV and a production cross section $\sigma (\app \to \att
+X)= 13.9^{+6.1}_{-4.8}~pb$ have been reported \cite{CDFmt}. The CDF value
for the top quark mass is in very comfortable agreement with the prediction
based on the SM electroweak fits of the LEP and SLC data, $\mt =177 \pm 11
^{+18}_{-19}$ GeV \cite{LEPew}. The top quark production cross section
measured by CDF is roughly a factor $\sim 2$ larger than the expected
theoretical value in QCD \cite{CDFmt} but is consistent with the upper
limit presented by the D0 collaboration: $\sigma (\app \to \att +X) <
13~pb$ (95\% C.L.) for a top quark mass of 180 GeV \cite{D0mt}. The neat
overlap between the estimates of $\mt$ based on the SM-electroweak analysis
and its direct measurement, together with the implied dominance of the
decay mode $t \to W^+ b$, is a resounding success of the standard model
\cite{GSW}.

It is well appreciated that the top quark plays a crucial role in the
phenomenology of the electroweak interactions, flavour mixing, rare decay
rates and CP violation. Therefore the new experimental input for $\mt$,
while still not very precise, should help in reducing the present
uncertainties on the parameters of the Cabibbo-Kobayashi-Maskawa (CKM)
quark mixing matrix \cite{CKM}. Conversely, the knowledge of $\mt$ can be
used to restrict the range of the relevant hadronic matrix elements, which
in turn should help in firming up SM-based predictions for rare decays and
CP asymmetries in a number of $K$- and $B$-hadron decays. The aim of this
article is to update the profile of the CKM matrix elements, in particular
the CKM unitarity triangle, taking into account all present measurements
and theoretical estimates of hadronic matrix elements, along with their
uncertainties. In doing this update, we also include the improvements
reported in a number of measurements of the lifetime, mixing ratio, and the
CKM matrix elements $\absvcb$ and $\vert V_{ub}/V_{cb} \vert$ from $B$
decays, measured by the ARGUS, CLEO, CDF and LEP experiments. The allowed
ranges for the CP-violating phases that will be measured in $B$ decays,
characterized by $\sin 2\beta$, $\sin 2\alpha$ and $\sin^2\gamma$, are also
presented. They can be measured directly through asymmetries in the decays
$\bdbarp \to J/\psi K_S$, $\bdbarp \to \pi^+ \pi^-$, and in  $\bsbarp\ \to
D_s^\pm K^\mp$, respectively. We also give the allowed domains for two of
the angles, $(\sin 2\alpha,\sin 2\beta)$ and the SM estimates for the
\bsbsbar\ mixing parameter, $\xs$.


\bigskip
\noindent
{\bf 2. An Update of the CKM Matrix}
\bigskip

In updating the CKM matrix elements, we make use of the Wolfenstein
parametrization \cite{Wolfenstein}, which follows from the observation that
the elements of this matrix exhibit a hierarchy in terms of $\lambda$, the
Cabibbo angle. In this parametrization the CKM matrix can be written
approximately as
\beq
V_{CKM} \simeq \left(\matrix{
    1-{1\over 2}\lambda^2 & \lambda
                       & A\lambda^3 \left( \rho - i\eta \right) \cr
  -\lambda & 1-{1\over 2}\lambda^2 - i A^2 \lambda^4 \eta & A\lambda^2 \cr
   A\lambda^3\left(1 - \rho - i \eta\right) & -A\lambda^2 & 1 \cr}\right)~.
\label{CKM}
\eeq

We shall restrict ourselves to specifying the main input, pointing out the
significant changes in the determination of the CKM parameters $\lambda$,
$A$, $\rho$, and $\eta$, since we presented our earlier fits
\cite{AL92}-\cite{ALIINT94}.

We recall that $\vert V_{us}\vert$ has been extracted with good accuracy
from $K\to\pi e\nu$ and hyperon decays \cite{PDG} to be
\beq
\vert V_{us}\vert=\lambda=0.2205\pm 0.0018~.
\eeq
This agrees quite well with the determination of $V_{ud}\simeq 1-{1\over
2}\lambda^2$ from $\beta$-decay:
\beq
\vert V_{ud}\vert=0.9744\pm 0.0010~.
\eeq

The parameter $A$ is related to the CKM matrix element $V_{cb}$, which can
be obtained from semileptonic decays of $B$ mesons. We shall restrict
ourselves to the methods based on the heavy-quark effective theory (HQET)
to calculate the exclusive and inclusive semileptonic decay rates. In the
heavy quark limit it has been observed that all hadronic form factors in
semileptonic decays can be expressed in terms of a single function, the
Isgur-Wise function \cite{Wisgur}. It has been shown that the HQET-based
method works best for $B\to D^*l\nu$ decays, since these decays are
unaffected by $1/m_b$ corrections \cite{Luke,Boyd,Neubert}. Furthermore,
the perturbative corrections calculated in HQET turn out to be small
\cite{Neubert}. This method has been used by both the ARGUS and CLEO
collaborations to determine $\absvcb$.

Using HQET, the differential decay rate in $B \to D^* \ell \nu_\ell$ is
\begin{eqnarray}
\frac{d\Gamma (B \to D^* \ell \bar{\nu})}{d\omega }
&=& \frac{G_F^2}{48 \pi^3} (m_B-m_{D^*})^2 m_{D^*}^3 \eta_{A}^2
  \sqrt{\omega^2-1} (\omega + 1)^2 \\ \nonumber
&~& ~~~~~~~~~~~~~~\times [ 1+ \frac{4 \omega}{\omega + 1}
 \frac{1-2\omega r + r^2}{(1-r)^2}] \absvcb ^2 \xi^2(\omega) ,
\label{bdstara1}
\end{eqnarray}
where $r=m_{D^*}/m_B$, $\omega=v\cdot v'$ ($v$ and $v'$ are the
four-velocities of the $B$ and $D^*$ meson, respectively), and $\eta_{A}$
is the short-distance correction to the axial vector form factor estimated
to be $\eta_{A}=0.99$ \cite{Neubert}. In the absence of any power
corrections $\xi (\omega=1)=1$. The size of the $O(1/\mb^2)$ and
$O(1/\mc^2)$ corrections to the Isgur-Wise function $\xi (\omega )$ has
recently become a matter of some discussion \cite{neuberttasi,suv94}. We
recall that the effects of such power corrections were previously estimated
as \cite{neuberttasi}:
\beq
\xi (1) = 1+ \delta (1/m^2)= 0.98 \pm 0.04
\label{neubertxi}
\eeq
In a recent paper Shifman, Uraltsev and Vainshtain \cite{suv94} have argued
that the deviation of $\xi (1)$ from unity is larger than the estimate
given in eq.~(\ref{neubertxi}). Following \cite{suv94}, this deviation can
be expressed as:
\beq
1-\xi^2(1) = \frac{1}{3}\frac{\mu_G^2}{m_c^2} + \frac{\mu_\pi^2-\mu_G^2}
{4} \big(\frac{1}{m_c^2} + \frac{1}{m_b^2} + \frac{2}{3 m_cm_b}\big)
+ \sum_{i=1,2,...} \xi^2_{B \to excit} ,
\label{suveq1}
\eeq
where the contribution to the higher excited states is indicated by the
last term, and $\mu_G^2$ and $\mu_\pi^2$ parametrize the matrix elements of
the chromomagnetic and kinetic energy operators, respectively. These have
been estimated to be:
\begin{eqnarray}
\mu_G^2 &=& \frac{3}{4} (M_{B^*}^2 - M_B^2) \simeq 0.35 ~\mbox{GeV}^2 ,
 \\ \nonumber
 \mu_\pi^2 &=&(0.54 \pm 0.12) ~\mbox{GeV}^2~,
\label{suveq2}
\end{eqnarray}
where the numbers for $\mu_\pi^2$ are based on QCD sum rules
\cite{ballbraun}. Using the central value for this quantity and ignoring
the contribution of the excited states, one gets
\beq
 \eta_A\xi(1)=0.92~.
\label{suveq3}
\eeq
The contribution of the higher states is positive definite. However, its
actual value can only be guessed at present. Shifman et.~al.\ estimate
\cite{suv94}:
\beq
\eta_A \xi(1)=0.89 \pm 0.3~.
\label{suveq4}
\eeq
The values of $\eta_A \xi(1)$ given in eqs.~(\ref{suveq3}) and
(\ref{suveq4}) are substantially smaller than the estimates given in
eq.~(\ref{neubertxi}) (with $\eta_A =0.99$) and used in previous
theoretical and experimental analyses. If the estimates by Shifman et.~al.\
are valid then one must conclude that the $O(1/m_Q^2)$ corrections to $\xi
(1)$ are not as innocuous as claimed previously! In the analysis for
$\absvcb$ presented here, we shall use the maximum allowed value for
$\eta_A \xi (1)$ in the estimate of eq.~(\ref{suveq4}), i.e.\ $\eta_A \xi
(1)=0.92$. Clearly, a better theoretical calculation for the excited states
is needed which might be forthcoming as the contribution of the inelastic
channels in semileptonic $B$ decays is measured more accurately.

Not only are theoretical estimates for $\xi (1)$ in a state of flux, so are
experimental numbers! The previously reported value for $\absvcb$ by the
ARGUS collaboration from the decays $B \to D^* + \ell \bar{\nu}$ using the
HQET formalism yielded a value $\absvcb =0.047 \pm 0.007$ with a
considerably higher value for the slope of the Isgur-Wise function, $\xi'
(1) \equiv -\rho^2$, in the range $1.9 < \rho^2 < 2.3$ \cite{argusvcb}. In
a recent analysis by ARGUS, significantly lower values for both $\absvcb$
and $\rho^2$ have been obtained, yielding \cite{Schroederpc}:
\begin{eqnarray}
\absvcb \big(\frac{\tau(B_d^0)}{1.53 ~\mbox{ps}}\big)^{1/2}
 &=& 0.039\pm 0.005~,
\nonumber \\
 \rho^2 &=& 1.08 \pm 0.12~,
\label{newarguscb}
\end{eqnarray}
where the value of $\absvcb$ corresponds to a linear extrapolation of the
Isgur-Wise function $\xi(\omega) = 1-\rho^2(\omega -1)$ and the error
quoted includes also that from the $B_d^0$ lifetime. The slope parameter in
eq.~(\ref{newarguscb}) is now in agreement with the theoretical bounds,
which suggest $\rho^2 \leq 1$ \cite{Bjorken,Voloshin}.

The numbers obtained by the CLEO collaboration from a similar method are
\cite{cleovcb}:
\begin{eqnarray}
\absvcb \big(\frac{\tau(B_d^0)}{1.5 ~\mbox{ps}}\big)^{1/2}
 &=& 0.039\pm 0.006,
\nonumber \\
&~& 1.0 \pm 0.04 < \rho^2 < 1.2 \pm 0.7 .
\label{newcleocb}
\end{eqnarray}
where no constraints on the slope at zero recoil are assumed. The two
results are in remarkable agreement!

Using the ARGUS and CLEO values of $\absvcb$ given in
eqs.~(\ref{newarguscb}) and (\ref{newcleocb}), renormalizing the Isgur-Wise
function to $\eta_A \xi (1)= 0.92$, and using the LEP value for the mean
$B$ lifetime, $\langle \tau_B \rangle = 1.54 \pm 0.03~ps$, the updated
values for $\absvcb$ can be expressed as\footnote{If the $B_d^0$ lifetime,
$\tau (B_d^0)= 1.52 \pm 0.11~ps$ is used instead, this does not change the
result significantly.}:
\beq
\absvcb \big(\frac{\tau(B_d^0)}{1.54 \mbox{ps}}\big)^{1/2}
 \big(\frac{\eta_A \xi (1)}{0.92}\big) = 0.041 \pm 0.006
\eeq
The corresponding analysis for the inclusive semileptonic $B$ decays
incorporating the power corrections given in eq.~(\ref{suveq2}) above has
also been undertaken by Shifman et.~al.\ \cite{suv94}. Updating the
$B$-lifetime, their estimate of $\absvcb$ can be expressed as:
\beq
\absvcb = 0.041 \big(\frac{1.54 ~\mbox{ps}}{\tau(B_d^0)}\big)^{1/2}
 \big(\frac{{\cal B}_{SL} (B)}{0.106}\big)^{1/2}
 \big(\frac{4.8 ~\mbox{GeV}}{\mb}\big)^{5}
\eeq
in which an error of $\pm 3 \%$, mostly due to the uncertainty in the $b$
quark pole mass, or, equivalently, on $(m_b - m_c)$, has been anticipated
\cite{suv94}. The error estimate notwithstanding, the HQET-based analyses
of the exclusive and inclusive semileptonic $B$ decays are in better
quantitative agreement than they have a right to be!

For the purposes of the fits which follow, we shall use the value of the
CKM parameter $A$ obtained from the above HQET-based methods:
\beq
A = 0.84 \pm 0.12~. \label{Avalue}
\eeq

The other two CKM parameters $\rho$ and $\eta$ are constrained by the
measurements of $\vert V_{ub}/V_{cb}\vert$, $\abseps$ (the CP-violating
parameter in the kaon system), $\xd$ (\bdbdbar\ mixing) and (in principle)
$\epsilon^\prime/\epsilon$ ($\Delta S=1$ CP-violation in the kaon system).
We shall not discuss the constraints from $\epsilon^\prime/\epsilon$, due
to the various experimental and theoretical uncertainties surrounding it at
present, but take up the rest in turn and present fits in which the allowed
region of $\rho$ and $\eta$ is shown.

First of all, $\vert V_{ub}/V_{cb}\vert$ can be obtained by looking at the
endpoint of the inclusive lepton spectrum in semileptonic $B$ decays. The
present average of this ratio, based on the recent analysis of the ARGUS
\cite{argusbu} and CLEO \cite{cleo15bu,cleoIIbu} data, is
\cite{ALIINT94,Stonebu}:
\beq
\left\vert \frac{V_{ub}}{V_{cb}} \right\vert = 0.08\pm 0.02~.
\label{vubvcbn}
\eeq
This gives
\beq
\sqrt{\rho^2 + \eta^2} = 0.36 \pm 0.09~.
\eeq

The experimental value of $\abseps$ is \cite{PDG}
\beq
\abseps = (2.26\pm 0.02)\times 10^{-3}~.
\eeq
Theoretically, $\abseps$ is essentially proportional to the imaginary part
of the box diagram for \kkbar\ mixing and is given by \cite{Burasetal}
\begin{eqnarray}
\abseps &=& \frac{G_F^2f_K^2M_KM_W^2}{6\sqrt{2}\pi^2\Delta M_K}
B_K\left(A^2\lambda^6\eta\right)
\bigl(y_c\left\{\eta_{ct}f_3(y_c,y_t)-\eta_{cc}\right\} \nonumber \\
&~& ~~~~~~~~~~~~~~~+ \eta_{tt}y_tf_2(y_t)A^2\lambda^4(1-\rho)\bigr).
\label{eps}
\end{eqnarray}
Here, the $\eta_i$ are QCD correction factors, $\eta_{cc}\simeq 0.82$,
$\eta_{tt}\simeq 0.62$, $\eta_{ct}\simeq 0.35$ for $\Lambda_{QCD}=200$ MeV
\cite{Flynn}, $y_i\equiv m_i^2/M_W^2$, and the functions $f_2$ and $f_3$
are given by
\begin{eqnarray}
f_2(x) &=& \frac{1}{4} + \frac{9}{4}\frac{1}{(1-x)}
- \frac{3}{2}\frac{1}{(1-x)^2}
- \frac{3}{2} \frac{x^2\ln x}{(1-x)^3}~, \nonumber \\
f_3(x,y) &=& \ln \frac{y}{x} - \frac{3y}{4(1-y)}
\left( 1 + \frac{y}{1-y}\ln y\right).
\end{eqnarray}
(The above form for $f_3(x,y)$ is an approximation, obtained in the limit
$x\ll y$. For the exact expression, see ref.~\cite{InamiLim}.)

The final parameter in the expression for $\abseps$ is $B_K$, which
represents our ignorance of the matrix element $\langle K^0 \vert
{({\overline{d}}\gamma^\mu (1-\gamma_5)s)}^2 \vert
{\overline{K^0}}\rangle$. The evaluation of this matrix element has been
the subject of much work. The results are summarized in ref.~\cite{AL92}.
Considering all the various calculational techniques, one is led to the
range $1/3 \le B_K \le 1$. However, the $1/N$ and lattice approaches are
generally considered the most reliable. They yield:
\beq
B_K = 0.8 \pm 0.2.
\label{BKrange}
\eeq

We now turn to \bdbdbar\ mixing. The latest value of $\xd$, which is a
measure of this mixing, is \cite{Danilov}
\beq
\xd = 0.71 \pm 0.07~.
\label{xdvalue}
\eeq
The mixing parameter $\xd$ is calculated from the \bdbdbar\ box diagram.
Unlike the kaon system, where the contributions of both the $c$- and the
$t$-quarks in the loop were important, this diagram is dominated by
$t$-quark exchange:
\beq
\label{bdmixing}
\xd \equiv \frac{\left(\Delta M\right)_B}{\Gamma}
= \tau_B \frac{G_F^2}{6\pi^2}M_W^2M_B\left(\fbb\right)\eta_By_t
f_2(y_t) \vert V_{td}^*V_{tb}\vert^2~, \label{xd}
\eeq
where, using eq.~\ref{CKM}, $\vert V_{td}^*V_{tb}\vert^2=
A^2\lambda^{6}\left[\left(1-\rho\right)^2+\eta^2\right]$. Here, $\eta_B$ is
the QCD correction. In ref.~\cite{etaB}, this correction is analyzed
including the effects of a heavy $t$-quark. It is found that $\eta_B$
depends sensitively on the definition of the $t$-quark mass, and that,
strictly speaking, only the product $\eta_B(y_t)f_2(y_t)$ is free of this
dependence. In the fits presented here we use the value $\eta_B=0.55$,
following ref.~\cite{etaB}.

For the $B$ system, the hadronic uncertainty is given by $\fbb$, analogous
to $B_K$ in the kaon system, except that in this case, also $\fbd$ is not
measured. Most lattice-QCD based estimates, as well as those from the QCD
sum rules, are compatible with the following ranges for $\fbb$ and
$B_{B_d}$ \cite{Sommer94,Narison}:
\begin{eqnarray}
\fbd &=& 180 \pm 50 ~\mbox{MeV}, \nonumber \\
B_{B_d} &=& 1.0 \pm 0.2
\label{FBrange}
\end{eqnarray}


\bigskip
\noindent
{\bf 3. The unitarity triangle}
\bigskip

The allowed region in $\rho$-$\eta$ space can be displayed quite elegantly
using the so-called unitarity triangle. The unitarity of  the CKM matrix
leads to the following relation:
\beq
V_{ud} V_{ub}^* + V_{cd} V_{cb}^* + V_{td} V_{tb}^* = 0~.
\eeq
Using the form of the CKM matrix in eq.~\ref{CKM}, this can be recast as
\beq
\frac{V_{ub}^*}{\lambda V_{cb}} + \frac{V_{td}}{\lambda V_{cb}} = 1~,
\eeq
which is a triangle relation in the complex plane (i.e.\ $\rho$-$\eta$
space), illustrated in Fig.~\ref{triangle}. Thus, allowed values of $\rho$
and $\eta$ translate into allowed shapes of the unitarity triangle.

\begin{figure}
\vspace{2.0 in}
\caption{The unitarity triangle. The angles $\alpha$, $\beta$ and $\gamma$
can be measured via CP violation in the $B$ system.}
\label{triangle}
\end{figure}

In order to find the allowed unitarity triangles, the computer program
MINUIT is used to fit the CKM parameters $A$, $\rho$ and $\eta$ to the
experimental values of $\absvcb$, $\vert V_{ub}/V_{cb}\vert$, $\abseps$ and
$\xd$. Since $\lambda$ is very well measured, we have fixed it to its
central value given above. The new ingredient in these fits is the value of
the top quark, for which we use the value $\mt=(174 \pm 16)$ GeV, measured
by CDF \cite{CDFmt}. We present here two types of fits:
\begin{itemize}
\item
Fit 1: the ``experimental fit.'' Here, only the experimentally measured
numbers are used as inputs to the fit with Gaussian errors; the coupling
constant $f_{B_d} \sqrt{B_{B_d}}$ and $B_K$ are given fixed values.
\item
Fit 2: the ``combined fit.'' Here, both the experimental and theoretical
numbers are used as inputs assuming Gaussian errors for the theoretical
quantities.
\end{itemize}

We first discuss the ``experimental fit" (Fit 1). The goal here is to
restrict the allowed ranges of the CKM parameters ($\rho,\eta)$ for given
values of the coupling constants, $f_{B_d} \sqrt{B_{B_d}}$ and $B_K$. We
first fix the bag factor to the value $B_K = 0.8$ and vary the value of the
coupling constant $\fbd\sqrt{B_{B_d}}$. The effect of varying the bag
factor $B_K$ in the range $B_K = 0.8 \pm 0.2$ is not crucial and will be
discussed later. The resulting fits for fixed values of
$\fbd\sqrt{B_{B_d}}$ are shown in Fig.~\ref{rhoeta1}. In all these graphs,
the solid line has $\chi^2 = \chi^2_{min} + 1$. Note that this corresponds
to only a 39\% confidence level region \cite{PDG}! For comparison, we
include the dashed line, which is the 90\% C.L.\ region ($\chi^2 =
\chi^2_{min} + 4.6$). As we pass from Fig.~\ref{rhoeta1}(a) to
Fig.~\ref{rhoeta1}(e), the most likely unitarity triangles become more and
more obtuse. This behaviour has already been anticipated
\cite{AL92,Martinelli,Schubert}. However, unlike the previous such
analyses, now that the top quark mass has been measured, the dependence of
the CKM triangle on the coupling constant product $\fbd\sqrt{B_{B_d}}$ can
be disentangled from that on $\mt$. As shown in these fits, the allowed
region in $(\rho,\eta)$-space is now quite restricted for a given value of
the coupling constant. This underscores the importance of measuring this
quantity, for example through the decays $B^\pm \to \tau^\pm \nu$.

The most probable values of the parameters $(\rho,\eta)$ are given in
Table
\ref{noxsfit}, together with their $\chi^2$. We remark that the values
$f_{B_d} \sqrt{B_{B_d}} \leq 110$ MeV and $f_{B_d} \sqrt{B_{B_d}} \geq 250$
MeV give very poor fits of the existing data. The minimum and maximum
allowed values of $f_{B_d} \sqrt{B_{B_d}}$ are somewhat correlated with the
value of $B_K$, which is the only remaining theoretical parameter of the
fit. For lower values of $B_K$, higher values of $f_{B_d} \sqrt{B_{B_d}}$
are disfavoured, while for higher values of $B_K$, somewhat higher values
of the coupling constant are allowed. Specifically, for $B_K =0.6$, the
allowed range is $110~\mbox{MeV} \leq f_{B_d} \sqrt{B_{B_d}} \leq
220~\mbox{MeV}$, whereas for $B_K=1.0$, the corresponding range is
$110~\mbox{MeV} \leq f_{B_d} \sqrt{B_{B_d}} \leq 270~\mbox{MeV}$. For the
lower value $B_K=0.4$, which is not favoured by the lattice and QCD sum
rules, the allowed range of $f_{B_d} \sqrt{B_{B_d}}$ is restricted to the
range 110-180 MeV, with generally higher values of $\chi^2$ than for the
case $B_K$ in the range 0.6-1.0. This suggests that present data disfavour
(though do not exclude) $B_K \leq 0.4$ solutions. Summing up, present data
exclude all values of $f_{B_d}\sqrt{f_{B_d}}$ which lie below 110 MeV and
above 270 MeV for the entire $B_K$ range.

\begin{table}
\hfil
\vbox{\offinterlineskip
\halign{&\vrule#&
   \strut\quad#\hfil\quad\cr
\noalign{\hrule}
height2pt&\omit&&\omit&&\omit&\cr
& $\fbd\sqrt{B_{B_d}}$ (MeV) && $(\rho,\eta)$ && $\chi^2_{min}$ & \cr
height2pt&\omit&&\omit&&\omit&\cr
\noalign{\hrule}
height2pt&\omit&&\omit&&\omit&\cr
&  $130$ && $(-0.33, ~0.18)$ && $0.10$ & \cr
&  $155$ && $(-0.27, ~0.27)$ && $0.14$ & \cr
&  $180$ && $(-0.05, ~0.33)$ && $0.33$ & \cr
&  $205$ && $(0.15, ~0.32)$ && $0.03$ & \cr
&  $230$ && $(0.28, ~0.30)$ && $0.39$ & \cr
height2pt&\omit&&\omit&&\omit&\cr
\noalign{\hrule}}}
\caption{The ``best values'' of the CKM parameters $(\rho,\eta)$ as a
function of the coupling constant $\fbd\protect\sqrt{B_{B_d}}$, obtained by
a minimum $\chi^2$ fit to the experimental data discussed in the text
including the CDF value $m_t=174 \pm 16$ GeV. We fix $B_K=0.8$. The
resulting minimum $\chi^2$ values from the MINUIT fits are also given.}
\label{noxsfit}
\end{table}

\begin{figure}
\vspace{7.75 in}
\caption{Allowed region in $\rho$-$\eta$ space, from a fit to the
experimental values given in the text, including $\mt=174 \pm 16$ GeV. We
have fixed $B_K=0.8$ and vary the coupling constant product
$\fbd\protect\sqrt{B_{B_d}}$ as indicated on the figures. The solid line
represents the region with $\chi^2=\chi_{min}^2+1$; the dashed line denotes
the 90\% C.L.\ region. The triangles show the best fit.}
\label{rhoeta1}
\end{figure}

One notices in Table \ref{noxsfit} that certain values of the coupling
constant $f_{B_d} \sqrt{B_{B_d}}$ give smaller $\chi^2$ than others. Indeed,
as one scans through the allowed parameter space in the coupling constants,
one obtains a double-valleyed solution corresponding to two minima in
$\chi^2$. For example, for $B_K=0.8$, there are minima in the $\chi^2$
distribution at $f_{B_d} \sqrt{B_{B_d}} = 140$ and 210 MeV. We do not
believe, however, that any exciting conclusions can be drawn from this
observation. The other values of $f_{B_d} \sqrt{B_{B_d}}$ in the range
110-250 MeV also give quite good fits to the data, so that the presence of
the minima at 140 and 210 MeV is not statistically significant.

\begin{figure}
\vspace{2.0 in}
\caption{Allowed region in $\rho$-$\eta$ space from a simultaneous fit to
both the experimental values given in the text (including $\mt=174 \pm 16$
GeV) and the theoretical quantities $B_K$ and $\fbd\protect\sqrt{B_{B_d}}$.
The theoretical errors are treated as Gaussian for this fit. The solid line
represents the region with $\chi^2=\chi_{min}^2+1$; the dashed line denotes
the 90\% C.L.\ region. The triangles show the best fit.}
\label{rhoeta2}
\end{figure}

We now discuss the ``combined fit" (Fit 2). Strictly speaking, this fit is
not on the same footing as the ``experimental fit" presented above, since
theoretical ``errors'' are not Gaussian. On the other hand, experimental
systematic errors are also not Gaussian, but it is common practice to treat
them as such, and to add them in quadrature with statistical errors. In
this sense, the method used in this fit is not unreasonable. Since the
coupling constants are not known and the best we have are estimates given
in the ranges in eqs.~(\ref{BKrange}) and (\ref{FBrange}), which are
allowed by data, a reasonable profile of the unitarity triangle at present
can be obtained by letting the coupling constants vary in this range. The
resulting CKM triangle is shown in Fig.~\ref{rhoeta2}, which still leaves
quite a bit of uncertainty in the $(\rho,\eta)$-space, though it is much
reduced compared to the previous such analyses, due to the knowledge of
$\mt$. The preferred values obtained from the ``combined fit" are
\beq
(\rho,\eta) = (0.14,0.32) ~~~(\mbox{with} ~\chi^2 =0.17).
\eeq
The resulting unitarity triangle is almost identical to the one for the
``experimental fit" (Fit 1) with $\fbd\sqrt{B_{B_d}}=205$ MeV, shown in
Fig.~\ref{rhoeta1}(d), which is suggestive but not compelling.


\bigskip
\noindent
{\bf 4. $\xs$ and the unitarity triangle}
\bigskip

Mixing in the \bsbsbar\ system is quite similar to that in the \bdbdbar\
system. The \bsbsbar\ box diagram is again dominated by $t$-quark exchange,
and the mixing parameter $\xs$ is given by a formula analogous to that of
eq.~(\ref{xd}):
\beq
\xs \equiv \frac{\left(\Delta M\right)_{B_s}}{\Gamma_{B_s}}
= \tau_{B_s} \frac{G_F^2}{6\pi^2}M_W^2M_{B_s}\left(\fbbs\right)
\eta_{B_s} y_t f_2(y_t) \vert V_{ts}^*V_{tb}\vert^2~. \label{xs}
\eeq
Using the fact that $\vert V_{cb}\vert=\vert V_{ts}\vert$ (eq.~\ref{CKM}),
it is clear that one of the sides of the unitarity triangle, $\vert
V_{td}/\lambda V_{cb}\vert$, can be obtained from the ratio of $\xd$ and
$\xs$:
\beq
\frac{\xs}{\xd} = \frac{\tau_{B_s}\eta_{B_s}M_{B_s}\left(\fbbs\right)}
{\tau_{B_d}\eta_{B_d}M_{B_d}\left(\fbb\right)}
\left\vert \frac{V_{ts}}{V_{td}} \right\vert^2.
\label{xratio}
\eeq
All dependence on the $t$-quark mass drops out, leaving the square of the
ratio of CKM matrix elements, multiplied by a factor which reflects
$SU(3)_{\rm flavour}$ breaking effects. The only real uncertainty in this
factor is the ratio of hadronic matrix elements. Whether or not $\xs$ can
be used to help constrain the unitarity triangle will depend crucially on
the theoretical status of the ratio $\fbbs/\fbb$.

The lifetime of the $B_s$ meson has now been measured at LEP and Tevatron.
The present average for the $B_s^0$ lifetimes is \cite{Sharma}:
\begin{equation}
\tau_{\bs} = (1.50 \pm 0.18) \times 10^{-12}~\mbox{s} ~.
\label{taubs}
\end{equation}
Within the experimental errors, this value is consistent with the averaged
value of $\tau_B$ used in the previous section. The mass of the $\bs$ meson
has also now been measured and its present best measurement is from CDF:
$\langle M_{B_s} \rangle = 5367.7 \pm 2.4 \pm 4.8$ MeV \cite{Muellermsm},
very close to the ALEPH measurement $\langle M_{B_s} \rangle = 5368.6 \pm
5.6 \pm 1.5$ MeV \cite{alephbsm}, with the DELPHI result $\langle M_{B_s}
\rangle = 5374.6 \pm 16 \pm 2$ MeV \cite{delphibsm} quite compatible with
the other two. We expect the QCD correction $\eta_{B_s}$ to be equal to its
$B_d$ counterpart, i.e.\ $\eta_{B_s} =0.55$. The main uncertainty in $\xs$
is now $\fbbs$. Using the determination of $A$ above, $\tau_{B_s}$ from
eq.~(\ref{taubs}) and $\mt =174 \pm 16$ GeV, we obtain
\begin{equation}
\xs = \left(428 \pm 147\right)\frac{\fbbs}{(1~\mbox{GeV})^2}.
\end{equation}
With $f_{B_s}\sqrt{B_{B_s}}= 170$ -- 300 MeV, this gives at 90\% C.L.\
(defined as 1.64$\sigma$),
\begin{equation}
5.4 \leq \xs\leq 60.2.
\end{equation}
The standard model therefore predicts very large values for $\xs$.

Another estimate can be obtained by using the relation in
eq.~(\ref{xratio}). Two quantities are required. First, we need the CKM
ratio $\vert V_{ts}/V_{td} \vert$. From our ``experimental fit," we have
obtained the 90\% C.L.\ bound on the inverse of this ratio
  as a function of
$f_{B_d}\sqrt{B_{B_d}}$. This is shown in Fig.~\ref{vtdts}. From this we
find
\beq
2.91 \leq \left\vert {V_{ts} \over V_{td}} \right\vert \leq 8.06
\eeq
The second ingredient is the SU(3)-breaking ratio
$f_{B_s}^2B_{B_s}/f_{B_d}^2B_{B_d}$. Estimating this to be 1.2 -- 1.5, we
determine at $90 \%$ C.L.:
\beq
7.2 \leq \xs \leq 52 .
\eeq
for $\xd =0.71$ \footnote{Folding in also the present experimental error on
$\xd$, the corresponding range is $6.1 \leq \xs \leq 79.9.$}. Thus, the two
ranges of $\xs$ are quite comparable and a value of $\xs \leq 5$ is
disfavoured in SM.

\begin{figure}
\vspace{2.5 in}
\caption{Allowed values of the CKM matrix element ratio $\vert
V_{td}/V_{ts} \vert$ as a function of the coupling constant product
$f_{B_d}\protect\sqrt{B_{B_d}}$, from the ``experimental fit" shown in
Fig.~\protect\ref{rhoeta1}. The solid line corresponds to the best fit
values and the dashed curves correspond to the maximum and minimum allowed
values at 90 \% C.L.}
\label{vtdts}
\end{figure}


\bigskip
\noindent
{\bf 5. CP Violation in the $B$ System}
\bigskip

It is expected that the $B$ system will exhibit large CP-violating effects,
characterized by nonzero values of the angles $\alpha$, $\beta$ and
$\gamma$ in the unitarity triangle (Fig.~\ref{triangle}) \cite{BCPasym}.
These angles can be measured via CP-violating asymmetries in hadronic $B$
decays. In the decays $\bdbarp \to \pi^+ \pi^-$, for example, one measures
the quantity $\sin 2\alpha$, and in $\bdbarp\to J/\psi K_S$, $\sin 2\beta$
is obtained. The CP asymmetry in the decay $\bsbarp\to D_s^\pm K^\mp$ is
slightly different, yielding $\sin^2 \gamma$.

These CP-violating asymmetries can be expressed straightforwardly in terms
of the CKM parameters $\rho$ and $\eta$. The 90\% C.L.\ constraints on
$\rho$ and $\eta$ found previously can be used to predict the ranges of
$\sin 2\alpha$, $\sin 2\beta$ and $\sin^2 \gamma$ allowed in the standard
model. The allowed ranges which correspond to each of the figures in
Fig.~\ref{rhoeta1}, obtained from the ``experimental fit" (Fit 1), are found
in Table \ref{CPranges}. In this Table we have assumed that the angle
$\beta$ is measured in $\bdbarp\to J/\Psi K_S$, and have therefore included
the extra minus sign due to the CP of the final state.

\begin{table}
\hfil
\vbox{\offinterlineskip
\halign{&\vrule#&
   \strut\quad#\hfil\quad\cr
\noalign{\hrule}
height2pt&\omit&&\omit&&\omit&&\omit&\cr
& $\fbd\sqrt{B_{B_d}}$ (MeV) && $\sin 2\alpha$ &&
$\sin 2\beta$ && $\sin^2 \gamma$ & \cr
height2pt&\omit&&\omit&&\omit&&\omit&\cr
\noalign{\hrule}
height2pt&\omit&&\omit&&\omit&&\omit&\cr
& $130$ && 0.43 - 0.98 && 0.18 - 0.41 && 0.1 - 0.56 & \cr
& $155$ && 0.42 - 1.0 && 0.28 - 0.62 && 0.29 - 1.0 & \cr
& $180$ && $-$1.0 - 1.0 && 0.33 - 0.78 && 0.2 - 1.0 & \cr
& $205$ && $-$1.0 - 0.91 && 0.38 - 0.87 && 0.14 - 1.0 & \cr
& $230$ && $-$1.0 - 0.53 && 0.46 - 0.92 && 0.12 - 0.97 & \cr
height2pt&\omit&&\omit&&\omit&&\omit&\cr
\noalign{\hrule}}}
\caption{The allowed ranges for the CP asymmetries $\sin 2\alpha$, $\sin
2\beta$ and $\sin^2 \gamma$, corresponding to the constraints on $\rho$ and
$\eta$ shown in Figs.~\protect\ref{rhoeta1}. Values of the coupling
constant $\fbd\protect\sqrt{B_{B_d}}$ are stated. The range for $\sin
2\beta$ includes an additional minus sign due to the CP of the final state
$J/\Psi K_S$.}
\label{CPranges}
\end{table}

Since the CP asymmetries all depend on $\rho$ and $\eta$, the ranges for
$\sin 2\alpha$, $\sin 2\beta$ and $\sin^2 \gamma$ shown in Table
\ref{CPranges} are correlated. That is, not all values in the ranges are
allowed simultaneously. We illustrate this in Fig.~\ref{alphabeta1},
corresponding to the ``experimental fit" (Fit 1), by showing the region in
$\sin 2\alpha$-$\sin 2\beta$ space allowed by the data, for various values
of $\fbd\sqrt{B_{B_d}}$. Given a value for $\fbd\sqrt{B_{B_d}}$, the CP
asymmetries are fairly constrained. However, since there is still
considerable uncertainty in the values of the coupling constants, a more
reliable profile of the CP asymmetries at present is given by our
``combined fit" (Fit 2), where we convolute the present theoretical and
experimental values in their allowed ranges. The resulting correlation is
shown in Fig.~\ref{alphabeta2}. From this figure one sees that the smallest
value of $\sin 2\beta$ occurs in a small region of parameter space
around $\sin 2\alpha\simeq 0.4$-0.6. Excluding this small tail, one expects
the CP-asymmetry in $\bdbarp\to J/\Psi K_S$ to be at least 30\%.

\begin{figure}
\vspace{8.0 in}
\caption{Allowed values of the CP asymmetries $\sin 2\alpha$ and $\sin
2\beta$ resulting from the ``experimental fit" of the data for different
values of the coupling constant $\fbd\protect\sqrt{B_{B_d}}$ indicated on
the figures a) -- e).}
\label{alphabeta1}
\end{figure}

\begin{figure}
\vspace{2.0 in}
\caption{Allowed values of the CP asymmetries $\sin 2\alpha$ and $\sin
2\beta$ resulting from the ``combined fit" of the data for the ranges for
$\fbd\protect\sqrt{B_{B_d}} $ and $B_K$ given in the text.}
\label{alphabeta2}
\end{figure}


\bigskip
\noindent
{\bf 6. Summary and Outlook}
\bigskip

We summarize our results:

\smallskip

(i) We have presented an update of the CKM unitarity triangle following
from the additional experimental input of $\mt=174 \pm 16$ GeV
\cite{CDFmt}. The fits can be used to exclude extreme values of the
pseudoscalar coupling constants, with the range $110~\mbox{MeV} \leq
f_{B_d} \sqrt{B_{B_d}} \leq 270~\mbox{MeV}$, still allowed for all values
of $B_K$. The solutions for $B_K=0.8\pm0.2$ are favoured by the data as
compared to the lower values. These numbers are in very comfortable
agreement with QCD-based estimates from sum rules and lattice techniques.
The statistical significance of the fit is, however, not good enough to
determine the coupling constant more precisely.

\smallskip

(ii) The allowed CKM unitarity triangle in the $(\rho,\eta)$-space is more
restricted than obtained previously without the top quark mass input.
However, the present uncertainties are still large unless the pseudoscalar
coupling constant could be determined independently. It may be possible to
measure the parameter $\fbd$, using isospin symmetry, via the
charged-current decay $\bu\to\tau^\pm \nu_\tau$. With $\vert V_{ub}/V_{cb}
\vert =0.08 \pm 0.02$ and $\fbd=180\pm 50~{\rm MeV}$, one gets a branching
ratio $BR(\bu\to\tau^\pm\nu_\tau)=(0.3$-$2.4)\times 10^{-4}$, which lies in
the range of the future LEP and asymmetric $B$-factory experiments. Along the
same lines, the prospects for measuring $(\fbd,\fbs)$ in the FCNC leptonic
and photonic decays of $\bd $ and $\bs$ hadrons, $(\bd,\bs)\to\mu^+\mu^-,
(\bd,\bs)\to\gamma\gamma$ in future $B$ physics facilities are not entirely
dismal \cite{ALIINT94}.

\smallskip

(iii) We have determined bounds on the ratio $\vert V_{td}/V_{ts} \vert$
from our fits. For $110~\mbox{MeV} \leq f_{B_d} \sqrt{B_{B_d}} \leq
270~\mbox{MeV}$, i.e.\ in the entire allowed domain, we find at 90 \% C.L.:
\beq
0.12 \leq \left\vert {V_{td} \over V_{ts}} \right\vert \leq 0.34~.
\eeq
The upper bound from our analysis is more restrictive than the current
experimental upper limit following from the CKM-suppressed radiative
penguin decays ${\cal B}(B \to \omega + \gamma )$ and ${\cal B}(B \to \rho
+ \gamma )$, which at present yield at 90\% C.L.\ \cite{cleotdul}:
\beq
\left\vert {V_{td} \over V_{ts}} \right\vert \leq \frac{1}{1.8}~.
\eeq
Furthermore, both the upper and lower bounds are better than those obtained
from unitarity, which gives $0.06 \leq \vert V_{td}/V_{ts} \vert \leq
0.59$ \cite{PDG}.

\smallskip

(iv) Using the measured value of $\mt$, we find
\begin{equation}
\xs = \left(428 \pm 147\right)\frac{\fbbs}{(1~\mbox{GeV})^2}.
\end{equation}
For$f_{B_s}\sqrt{B_{B_s}}= 170$ -- 300 MeV, this gives
\begin{equation}
5.4 \leq \xs\leq 60.2
\end{equation}
at 90\% C.L.

\smallskip

(v) The ranges for the CP-violating asymmetries $\sin 2\alpha$, $\sin
2\beta$ and and $\sin^2 \gamma$ are determined to be at 90\% C.L.:
\begin{eqnarray}
&~& -1.0 \leq \sin 2\alpha \le 1.0~, \nonumber \\
&~& 0.16 \leq \sin 2\beta \le 0.91~, \\
&~& 0.09 \leq \sin^2 \gamma \le 1.0~. \nonumber
\end{eqnarray}
(For $\sin 2\alpha < 0.4$, we find $\sin 2\beta \ge 0.3$.)

\bigskip
\noindent
{\bf Acknowledgements}:
\bigskip

We thank Henning Schr\"oder and Vivek Sharma for providing updated analysis
of $\absvcb$ and $B$-lifetimes, respectively. We thank Christoph Greub,
Thomas Mannel and Nicolai Uraltsev for discussions.



\end{document}